# The Impact of LLM Self-Consistency and Reasoning Effort on Automated Scoring Accuracy and Cost

**Scott Frohn**

*Khan Academy, Mountain View, CA, USA*
scottfrohn@khanacademy.org

**Abstract**

Strategic model selection and reasoning settings are more effective than ensembling for optimizing automated scoring with large language models (LLMs). We examined self-consistency (intra-model majority voting) and reasoning effort for scoring conversation-based assessment items in high school mathematics, evaluating 900 student conversations against human-scored ground truths using frontier and low-cost models from OpenAI and Google. Temperature sampling significantly improved accuracy over deterministic calls, but increasing ensemble size ($j = 1$ to $7$) produced no significant gains. Higher reasoning effort showed a significant positive linear trend with scoring accuracy, though the benefit varied by model family. An efficiency frontier analysis identified Gemini 3.1 Pro Preview at low reasoning as the most accurate but costly configuration; GPT-5.4 Nano and Mini with no reasoning offered the best cost-performance balance.

## 1 Introduction

Automated scoring of constructed response questions has long been a target of educational technology research. The core challenge is practical: scoring open-ended student responses at scale requires time and consistency that is difficult to achieve with human raters alone. Unlike multiple-choice items, short-answer questions require students to respond in their own words, making them better measures of conceptual understanding—but also far more costly to score. Early automated essay and short answer grading systems relied on surface-level linguistic features and traditional machine learning classifiers. They demonstrated modest agreement with human raters but required extensive labeled training data and could be brittle in the face of diverse student language (Burrows et al., 2015).

Large language models (LLMs) have expanded what automated scoring systems can do. Commercial LLMs – like those from OpenAI and Google – can evaluate student responses in a zero-shot (without scored examples) or few-shot (with scored examples) setting, requiring little or no fine-tuning. They condition on a rubric and return a score with a reasoning trace. Commercial LLM-based scoring shows promise (Frohn et al., 2025; Henkel et al., 2024; Lee et al., 2024; Pack et al., 2024; Xue et al., 2026), though performance depends on many factors, including model selection, model settings (i.e., temperature), prompt design, number of score levels, and rubric design.

One method for improving performance is *self-consistency* (Wang et al., 2023), which samples a model multiple times and aggregates the results via majority vote. Originally proposed to improve chain-of-thought reasoning on tasks with determinate correct answers, self-consistency has demonstrated gains of 4–18% on arithmetic and commonsense benchmarks (Wang et al., 2023). The appeal for scoring is straightforward: if individual model calls are noisy, aggregating across multiple independent calls should smooth out that noise and converge on a more reliable judgment. Varying temperature—the parameter that controls output randomness—could further diversify the sample and increase the odds of capturing the correct score. Some of our recent AI scoring research (Frohn et al., 2025) employed self-consistency with three judges ($j = 3$) and found high performance across multiple LLMs for dichotomously scored high school mathematics and English language arts (ELA) short answer questions. We did not, however, compare self-consistency against a single judge or other ensemble sizes.

Recent empirical work complicates this expectation. Xue et al. (2026) explored the performance of five models (GPT-4o, DeepSeek-V3, Gemini-2.0-flash, Claude-3.7-Sonnet, and Qwen-3-plus) representing a range of model architectures. Intra-model majority voting ($j = 5$) produced less than 1% improvement in quadratic weighted kappa (QWK) over a single call when scoring short response items on 2–4 point rubrics—a result that was not statistically significant. The problem appears to be a decoupling between consistency and accuracy: individual models score with near-perfect self-consistency (especially at temp = 0), but that consistency reflects a stable scoring pattern, correct or not. More sampling does not correct a model that is systematically misapplying a rubric. The authors also found that ensembling five different models improved performance above intra-model majority voting, but accuracy remained modest (average QWK $\leq 0.60$). These findings suggest that the theoretical basis for self-consistency—multiple reasoning paths converging on a unique correct answer—does not straightforwardly transfer to rubric-based scoring, where correctness is a matter of judgment rather than logical derivation.

Another method for improving performance is using model *reasoning level*, enabled by a new generation of models that generate internal chain-of-thought traces before producing a final response (Wei et al., 2023). Parameters such as OpenAI's `reasoning_effort` and Google's `thinking_level` control how deeply a model deliberates before scoring. Higher effort settings have shown substantial accuracy gains on complex mathematical and scientific reasoning tasks (OpenAI, 2024), but their value for automated scoring is less clear. Reasoning models can over-interpret student responses. Yoshida (2025) found that two of OpenAI's reasoning models (o3-mini and o4-mini) did not perform as well as the non-reasoning GPT-4o mini at TOEFL11 essay scoring, partly because LLMs tended to give higher scores than experts. Reasoning models also introduce latency and cost (Sui et al., 2026) that may be difficult to justify for large-scale educational assessments.

Self-consistency and higher reasoning effort may improve performance on benchmark tasks, but these methods have not yet reliably improved automated scoring in educational contexts. Prior investigations focused primarily on polytomously scored essay or short response questions in science and English-related subjects. Both methods need further exploration across different assessment contexts, subjects, rubrics, and item types.

Khan Academy has developed an "Explain Your Thinking" (EYT) item type for its interim assessments (DiCerbo, 2025). Following a selected response item, this conversation-based assessment item is guided by an AI assistant to further explore students' understanding. The AI assistant uses a checklist-style rubric of 2–3 dichotomously scored criteria to guide the conversation and evaluate the student's response. This context is well suited to extend existing research on self-consistency and model reasoning.

Both methods carry a monetary cost. Self-consistency increases scoring cost linearly with ensemble size: three judges cost roughly three times as much as one. How reasoning effort translates to cost is less clear. Commercial LLM providers typically disclose thinking tokens used by reasoning models, and these tokens are often billed at the output token rate (Google, 2026; OpenAI, 2026), but the amount of additional tokens varies. Any implementation using reasoning settings must therefore weigh higher reasoning against additional cost.

The present study investigates self-consistency and reasoning effort for EYT items on high school mathematics interim assessments. Using student conversations scored by trained human raters, we examine three research questions.

**R1:** Does self-consistency (intra-LLM ensembling) improve scoring accuracy?

Wang et al. (2023) suggest self-consistency may improve scoring accuracy, but recent empirical work in automated scoring contexts (Xue et al., 2026) gives reason to temper that expectation. The theoretical argument holds that multiple reasoning paths should converge on a more accurate solution (i.e., score). There is, however, a fundamental difference between solving reasoning tasks with an objective solution and evaluating student responses against a rubric, which requires subjective interpretation.

**R2:** Does increasing reasoning effort improve scoring accuracy, and does the benefit vary across model family and model size?

Yoshida (2025) found that OpenAI's reasoning models o3-mini and o4-mini showed only moderate agreement with humans when scoring TOEFL essays on a 3-point scale (QWKs $< 0.55$). Those are smaller models optimized for speed and cost-efficiency (OpenAI, 2025a, 2025b), and newer models—both frontier and lower-cost variants—make reasoning level a modifiable parameter. Manipulating reasoning level may therefore affect performance, and we expect higher reasoning levels to produce higher scoring accuracy.

**R3:** What is the optimal balance between performance and cost?

This is an exploratory question intended to provide practical guidance on where resources are best applied. The optimal balance between performance and cost depends on scoring accuracy—which itself depends on the nuances of the scoring task, model, and model settings—and pricing structure.

## 2 Method

### 2.1 Large Language Models

We selected two model families—OpenAI and Google—each offering a frontier model and lower-cost variants, adjustable temperature, and broadly available APIs. From OpenAI: the frontier model GPT-5.4 (released March 5, 2026) and the low-cost variants

GPT-5.4-Mini and GPT-5.4-Nano (both released March 17, 2026). From Google: the frontier Gemini 3.1 Pro Preview (released February 19, 2026) and the low-cost variant Gemini 3 Flash Preview (released December 17, 2025). All models were accessed via the OpenAI and Google APIs in R using the `ellmer` package, v.0.3.2 (Wickham et al., 2025).

## 2.2 Data

Three EYT items from high school mathematics were included: two from Algebra I and one from Geometry. Each item has two parts. Part 1 is a selected response item; Part 2 is a follow-up conversation in which an AI assistant probes the student's reasoning about Part 1. Each conversation is scored against a checklist rubric with two or three discrete criteria.

The conversation ends when (a) all criteria are satisfied, (b) the student provides four responses (the turn limit), or (c) the student exits early. Each criterion includes an indicator against which responses are judged, evaluation notes for interpreting responses, and examples of correct and incomplete responses.

We randomly sampled 300 student conversations per item (900 total). Three trained human raters scored each conversation on each criterion; majority vote across the three raters served as ground truth (2 × 300 + 3 × 300 + 2 × 300 = 2,100 ground truth scores). Table 1 presents item descriptives and Table 2 presents human interrater reliability.

**Table 1.** EYT Item Descriptive Information

| Item | Subject | Criteria | Students | Avg. Responses per Conversation | Avg. Conversation Length (Words) |
|---|---|---|---|---|---|
| Item 1 | Algebra I | 2 | 300 | 2.3 | 21.7 |
| Item 2 | Algebra I | 3 | 300 | 2.2 | 14.5 |
| Item 3 | Geometry | 2 | 300 | 2.6 | 22.0 |

**Table 2.** EYT Criterion-Level Human–Human Interrater Reliability

| Item | Criterion | Fleiss' $\kappa$ |
|---|---|---|
| Item 1 | 1 | 0.858 |
|  | 2 | 0.713 |
| Item 2 | 1 | 0.797 |
|  | 2 | 0.662 |
|  | 3 | 0.749 |
| Item 3 | 1 | 0.800 |
|  | 2 | 0.910 |

## 2.3 Prompt

Khan Academy developed the scoring prompt through iterative prompt engineering. The prompt scores one criterion per call; a conversation with three criteria requires three separate calls. The template is shown in Figure A1 in the Appendix.

## 3 Results

### 3.1 Self-Consistency

For R1, we restricted the self-consistency analysis to models where temperature could be meaningfully adjusted: the three GPT variants and Gemini 3 Flash Preview. GPT models support temperature adjustment only when `reasoning_effort = "none"`, so all three GPT variants were run at that setting. For Gemini, only the Flash variant at its lowest thinking level ("minimal") was included; Gemini 3.1 Pro Preview does not support a minimal thinking level, and Google cautions against modifying temperature for Gemini 3 models (Google, 2026). We excluded models at higher reasoning levels to avoid the cost and latency of repeated high-effort calls.

Intra-model consistency—how often a model agreed with itself across repeated calls—was high for all models and ensemble sizes: Fleiss' $\kappa$ ranged from 0.886 to 0.969 (Table 3). We computed this at the criterion level using the $j$ replications at temp = 1 for $j$ = 3, 5, and 7. This measure is analogous to interrater reliability. The result undermines the theoretical premise of self-consistency: individual calls are too consistent to benefit meaningfully from aggregation.

**Table 3.** Intra-model Consistency (Fleiss' $\kappa$) by LLM and Ensemble Size at Lowest-Effort Reasoning (temp = 1)

| Model | Reasoning | $j = 3$ | $j = 5$ | $j = 7$ |
|---|---|---|---|---|
| GPT-5.4 Nano | none | 0.886 | 0.890 | 0.887 |
| GPT-5.4 Mini | none | 0.908 | 0.907 | 0.903 |
| GPT-5.4 | none | 0.942 | 0.942 | 0.942 |
| Gemini 3 Flash Preview | minimal | 0.965 | 0.969 | 0.966 |

Because criteria are dichotomously scored, we used Cohen's $\kappa$ as the primary measure of agreement with human ground truth. We also scored at temp = 0 ($j = 1$) as a single-call baseline representative of common deployment practice. At $j = 1$, $\kappa$ ranged from 0.56 to 0.75 across models and temperature conditions, indicating moderate to substantial agreement with human raters (Table 4).

**Table 4.** Model Performance (Cohen's $\kappa$) by LLM and Ensemble Size at Lowest-Effort Reasoning

| | | temp = 0 | temp = 1 | | | |
|---|---|---|---|---|---|---|
| Model | Reasoning | $j = 1$ | $j = 1$ | $j = 3$ | $j = 5$ | $j = 7$ |
| GPT-5.4 Nano | none | 0.564 | 0.750 | 0.757 | 0.761 | 0.761 |
| GPT-5.4 Mini | none | 0.668 | 0.756 | 0.772 | 0.766 | 0.767 |
| GPT-5.4 | none | 0.708 | 0.701 | 0.714 | 0.714 | 0.714 |
| Gemini 3 Flash Preview | minimal | 0.714 | 0.713 | 0.717 | 0.717 | 0.720 |

We fit two generalized linear mixed models (GLMMs) using the `lme4` package in R (v.1.1-38; Bates et al., 2015). The GLMM preserves the observation-level variance structure that aggregate $\kappa$ discards. The outcome was binary accuracy (correct vs. incorrect LLM prediction per criterion per conversation).

The first GLMM compared a single call at temp = 0 to one at temp = 1 ($j = 1$ in both cases), with temperature as a fixed effect and random intercepts for item-criterion and model. Stochastic calls (temp = 1) were significantly more accurate than deterministic calls ($b = 0.193$, $z = 3.95$, $p < .001$). The effect was inconsistent across models: GPT-5.4 and Gemini 3 Flash Preview showed negligible temperature differences, while GPT-5.4 Mini and Nano gained more at temp = 1. The full model is presented in Table A1 in the Appendix.

The second GLMM tested majority-vote ensembling by comparing $j = 1$, 3, 5, and 7 at temp = 1, with ensemble size as a continuous fixed effect and random intercepts for item-criterion and model. Increasing ensemble size from $j = 1$ to $j = 7$ produced no significant accuracy gain ($b = 0.007$, $z = 0.88$, $p = .379$); $\kappa$ was flat across ensemble sizes for all four models (maximum gain $\leq 0.02$ points; Table A2 in the Appendix).

The high intra-model consistency observed here reflects stable scoring patterns—correct or not—rather than the variable reasoning that self-consistency aggregation is designed to exploit. Models performed at least as well—and often better—at temp = 1 than at temp = 0, and ensembling produced no meaningful accuracy gains. All remaining analyses are therefore based on a single judgment from each model at temp = 1.

### 3.2 Reasoning Level

Table 5 shows substantial heterogeneity in Cohen's $\kappa$ across models and reasoning levels. Performance plateaued at higher reasoning levels for most models. GPT-5.4 Nano and GPT-5.4 Mini performed better with no reasoning than with low reasoning.

**Table 5.** Cohen's $\kappa$ by Model and Reasoning Level ($j = 1$, temp = 1)

| Model | Reasoning Level | | | |
|---|---|---|---|---|
| | none/minimal | low | medium | high |
| GPT-5.4 Nano | 0.750 | 0.700 | 0.743 | 0.742 |
| GPT-5.4 Mini | 0.756 | 0.705 | 0.752 | 0.763 |
| GPT-5.4 | 0.701 | 0.760 | 0.759 | 0.763 |
| Gemini 3 Flash Preview | 0.713 | 0.710 | 0.782 | 0.771 |
| Gemini 3.1 Pro Preview | — | 0.794 | 0.788 | 0.793 |

For R2, we fit a single GLMM pooling both GPT and Gemini conditions (Table A3 in the Appendix), with reasoning level as an ordered factor (none/low/medium/high), model as an unordered fixed effect, and a random intercept for item-criterion. We recoded

Gemini's minimum level ("minimal") to "none" to align with GPT's lowest-effort setting—both represent the minimum reasoning available from each provider.

Higher reasoning effort significantly improved scoring accuracy (linear contrast: $b = 0.145$, $z = 4.06$, $p < .001$). Ordered factors in `lme4` are decomposed into orthogonal polynomial contrasts; the linear term is our primary test of monotonic improvement with reasoning effort. The quadratic contrast was not significant ($b = 0.008$, $z = 0.22$, $p = .823$), but the cubic contrast was marginally significant ($b = -0.069$, $z = -2.04$, $p = .041$), indicating departures from a strictly monotonic trend. No adjacent pairwise contrast reached significance after Bonferroni correction.

Gemini 3.1 Pro Preview significantly outperformed all other models ($b = 0.260$, $z = 4.33$, $p < .001$); its $\kappa$ ranged narrowly from 0.788 to 0.794 across reasoning levels, indicating little practical sensitivity to this parameter. The three GPT variants and Gemini 3 Flash Preview did not differ significantly from one another (all $p$s = 1.00 after Bonferroni correction). Within the GPT family, the reasoning–accuracy relationship was inconsistent: GPT-5.4 gained substantially from none to low reasoning ($\kappa = 0.701$ to 0.760) with little further improvement, while GPT-5.4 Mini and Nano dipped at low reasoning before recovering at medium and high. Gemini 3 Flash Preview showed the clearest trend, rising from $\kappa = 0.713$ at minimal reasoning to 0.782 at medium before leveling off at high.

### 3.3 Cost Efficiency

We estimated API cost by randomly sampling 20 conversations across items and criteria and submitting each to the model via a live API call. Token counts were extracted directly from the raw API response for each call, averaged across the 20 sampled conversations, and multiplied by standard (non-batch) API list prices per million tokens as of March 2026 (Google, 2026; OpenAI, 2026) to produce an estimated cost per call. All cost estimates reflect on-demand pricing without batch discounts. API costs are presented in Table 6.

**Table 6.** API Cost per Model[1]

| Model | Input Price | Output Price |
| --- | --- | --- |
| GPT-5.4 Nano | $0.20 | $1.25 |
| GPT-5.4 Mini | $0.75 | $4.50 |
| GPT-5.4 | $2.50 | $15.00 |
| Gemini 3 Flash Preview | $0.50 | $3.00 |
| Gemini 3.1 Pro Preview | $2.00 | $12.00 |

Input token counts are determined by the system prompt and conversation text, which are identical across conditions for the same criterion. We therefore used a single pooled average input length (1,030 tokens) across all conditions to reduce sampling noise. Output token counts were averaged within each condition, since output length varies by model and reasoning level.

---

[1] **Note.** *Costs reflect standard (non-batch) API pricing per million tokens as of March 2026 (Google, 2026; OpenAI, 2026).*

**Table 7.** Estimated API Cost per 1,000 Calls by Model and Reasoning Level[2]

| Model | Reasoning | Avg. Input Tokens | Avg. Output Tokens | Avg. Reasoning Tokens | Billed Output Tokens | Cost / 1k Calls |
|---|---|---|---|---|---|---|
| GPT-5.4 Nano | none | 1030 | 51 | 0 | 51 | $0.27 |
|  | low | 1030 | 126 | 70 | 126 | $0.36 |
|  | medium | 1030 | 176 | 119 | 176 | $0.43 |
|  | high | 1030 | 240 | 181 | 240 | $0.51 |
| GPT-5.4 Mini | none | 1030 | 40 | 0 | 40 | $0.95 |
|  | low | 1030 | 84 | 40 | 84 | $1.15 |
|  | medium | 1030 | 154 | 108 | 154 | $1.47 |
|  | high | 1030 | 241 | 194 | 241 | $1.85 |
| GPT-5.4 | none | 1030 | 43 | 0 | 43 | $3.22 |
|  | low | 1030 | 92 | 42 | 92 | $3.95 |
|  | medium | 1030 | 147 | 96 | 147 | $4.79 |
|  | high | 1030 | 236 | 188 | 236 | $6.12 |
| Gemini 3 Flash Preview | minimal | 1030 | 50 | 0 | 50 | $0.67 |
|  | low | 1030 | 46 | 0 | 46 | $0.65 |
|  | medium | 1030 | 50 | 716 | 766 | $2.81 |
|  | high | 1030 | 49 | 417 | 466 | $1.91 |
| Gemini 3.1 Pro Preview | low | 1030 | 52 | 100 | 152 | $3.88 |
|  | medium | 1030 | 50 | 120 | 170 | $4.11 |
|  | high | 1030 | 48 | 124 | 173 | $4.13 |

Table 7 presents estimated costs per 1,000 calls. Costs ranged from $0.27 for GPT-5.4 Nano with no reasoning to $6.12 for GPT-5.4 at high reasoning. Gemini 3 Flash Preview shows a counterintuitive pattern: medium reasoning averaged 716 thinking tokens per call versus 417 at high, driving its medium cost ($2.81) above its high cost ($1.91). We replicated this pattern across repeated sampling runs. Gemini 3.1 Pro Preview ranged from $3.88 to $4.13 per 1,000 calls—nearly flat across reasoning levels, as thinking token counts were stable once the model's deliberation was engaged.

---

[2] Costs are estimated based on average token usage across the 2,100 scored item-criterion-conversation observations at j = 1, temp = 1, and reflect standard (non-batch) API pricing as of March 2026 (Google, 2026; OpenAI, 2026). For OpenAI models, reasoning tokens are included within the reported output token count and billed at the output token rate. For Gemini models, reasoning and output tokens are reported separately but both billed at the output token rate; billed output tokens therefore reflect the sum of output and reasoning tokens.

To address R3, we constructed an efficiency frontier across all single-call conditions ($j = 1$, temp = 1), plotting each condition in $\kappa$-by-cost space and identifying Pareto-optimal points—conditions not dominated by any other with both higher $\kappa$ and lower cost. The frontier identifies the best accuracy achievable at each cost level, providing practical guidance for deployment decisions.

Figure 2 plots all single-call conditions in $\kappa$-by-cost space, with Pareto-optimal conditions connected by the dashed frontier line. GPT-5.4 Nano and GPT-5.4 Mini with no reasoning effort anchor the low-cost end of the frontier ($\kappa = 0.750$–$0.756$; $0.27 and $0.95 per 1,000 calls). Moving up the frontier, Gemini 3 Flash Preview at high reasoning ($1.91) offers a modest accuracy gain ($\kappa = 0.771$), followed by medium reasoning ($2.81, $\kappa = 0.782$). Gemini 3.1 Pro Preview at low reasoning anchors the high-performance end ($\kappa = 0.794$, $3.88 per 1,000 calls), delivering the strongest accuracy among all conditions.

Higher reasoning effort for GPT-5.4 and mid-tier GPT models fell off the frontier—it added cost without commensurate accuracy gains. Practitioners prioritizing cost efficiency can achieve near-frontier accuracy with GPT-5.4 Nano or Mini at no reasoning effort. Those prioritizing accuracy should consider Gemini 3.1 Pro Preview at low reasoning ($\kappa = 0.794$), though at four to fifteen times the cost of the most efficient configurations.

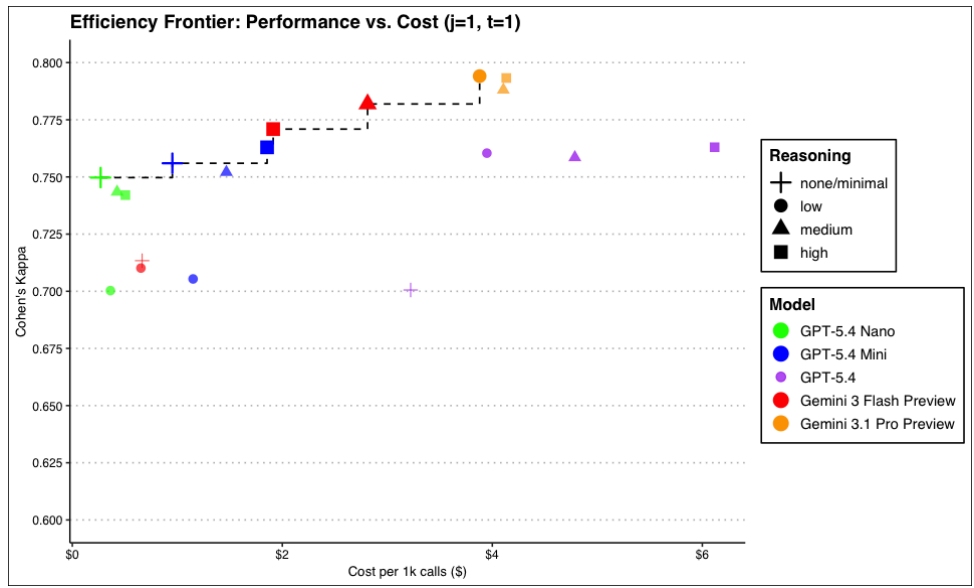

**Fig. 2** Efficiency frontier for LLM performance by model and reasoning level. Each point represents a single-call condition ($j = 1$, temp = 1). Pareto-optimal conditions are connected by the dashed frontier line.

## 4 Discussion

Increasing ensemble size from $j = 1$ to $j = 7$ at temp = 1 produced no significant accuracy gains (R1). This contradicts the theoretical premise established for reasoning tasks with deterministic answers, but aligns with prior empirical work (Xue et al., 2026) showing that intra-model consistency often decouples from actual scoring accuracy. Temperature manipulation (temp = 1 vs. temp = 0) improved accuracy for certain models; ensembling did not.

Higher reasoning effort improved scoring accuracy overall (R2), with a significant positive linear trend. The relationship was not strictly monotonic, however, and varied substantially across model families. Gemini 3.1 Pro Preview maintained superior performance regardless of reasoning level. Gemini 3 Flash Preview showed the most distinct sensitivity to changes in reasoning effort.

The efficiency frontier identified the practical tradeoffs for deployment (R3). At standard API pricing, Gemini 3.1 Pro Preview at low reasoning effort anchors the high-performance end of the frontier ($\kappa \approx 0.794$ at \$3.88 per 1,000 calls), while GPT-5.4 Nano and Mini with no reasoning effort deliver near-frontier performance at a fraction of the cost (\$0.27–\$0.95 per 1,000 calls). Whether Gemini 3.1 Pro Preview's accuracy advantage justifies its cost depends on the deployment context and the consequences of scoring errors. Higher reasoning effort did not consistently improve accuracy but did consistently increase cost; practitioners should evaluate reasoning levels rather than accept model defaults.

For deployments at scale, costs can be reduced further: batch processing APIs from both OpenAI and Google provide roughly 50% cost reductions relative to standard pricing; prompt caching can reduce input token costs for prompts that share a common prefix, such as a fixed rubric; and lower-cost model variants with minimal reasoning effort may be sufficient for many scoring tasks given the modest performance differences observed across model sizes. Thoughtful model and configuration selection—rather than defaulting to the most capable or highest-effort option—is likely to yield the best balance of accuracy and cost in operational automated short answer grading (ASAG) deployments.

## 5 Limitations

Several factors limit the generalizability of these findings. The study focused exclusively on dichotomously scored, checklist-style conversation-based items for high school mathematics; results may not generalize to polytomous scoring, essays, or other subject domains. We also restricted the self-consistency analysis to intra-LLM ensembling and did not explore inter-model ensembling across different architectures.

Cost estimates reflect standard (non-batch) API list prices as of March 2026 and do not account for batch discounts, prompt caching, or open-source model hosting—all of which would reduce costs substantially. Gemini 3 Flash Preview showed an anomalous cost pattern: medium reasoning generated more thinking tokens than high reasoning (716 vs. 417 on average), making medium more expensive. This pattern was consistent across repeated sampling runs and may reflect non-linear behavior at intermediate thinking levels.

Finally, this study is cross-sectional. It does not address whether LLM scoring accuracy and consistency remain stable as model providers update APIs. Given the pace of model development, the specific cost and performance figures reported here will shift over time and should be treated as point-in-time estimates.

**Appendix**

```
SYSTEM_PROMPT = """\
<Role>
You are an evaluator reviewing a conversation between an AI assessment
proctor and a student. Determine whether the student satisfied the
evaluation criterion based on their responses. Do NOT continue the
conversation. Do NOT roleplay as the proctor or the student.
</Role>
<Problem>{problem}</Problem>
<StudentAnswer>{student_answer}</StudentAnswer>
<Criterion>{criterion}</Criterion>
<Conversation>
{conversation}
</Conversation>
<Requirements>
    - Do NOT mark the criterion as satisfied unless you are confident
that the student has demonstrated understanding of the criterion.
    - Students do not need to use the EXACT concept terms in the
criterion, but can use synonymous language.
    - Evaluate based on the full context: the problem, the student's
answer, and their conversation.
    - Do not assume the student understands the criterion based on a
correct answer to the Problem.
    - Don't mark the criterion as satisfied if the student is
describing a step in solving the problem (unless that is necessary to
demonstrate understanding of the criterion).
</Requirements>
Respond with a JSON object matching this schema:
{
  "1_Reasoning": "Brief reasoning about whether the criterion is
satisfied (25 words max).",
  "2_IsSatisfied": false
}
- 1_Reasoning (string, required): Brief reasoning. Keep to 25 words at
most.
- 2_IsSatisfied (boolean, required): true if satisfied, false
otherwise.
Return ONLY the JSON object, no other text.
"""
```

**Fig. 1** Scoring prompt template used for all LLM calls. The template accepts `{problem}`, `{student_answer}`, `{criterion}`, and `{conversation}` as placeholders.

**Table A1.** Generalized Linear Mixed Model Predicting Scoring Accuracy from Temperature

|  | *b* | *SE* | *z* | *p* |
|---|---|---|---|---|
| **Fixed effects** | | | | |
| Intercept | 2.023 | 0.176 | 11.52 | < .001 |
| Temperature | 0.193 | 0.049 | 3.95 | < .001 |
| **Random effects** | | | | |
|  | *Variance* | *SD* | | |
| Item-criterion | 0.208 | 0.456 | | |
| Model | 0.000 | 0.000 | | |

**Table A2.** Generalized Linear Mixed Model Predicting Scoring Accuracy from Ensemble Size

|  | *b* | *SE* | *z* | *p* |
|---|---|---|---|---|
| **Fixed effects** | | | | |
| Intercept | 2.261 | 0.231 | 9.78 | < .001 |
| Ensemble size ($j$) | 0.007 | 0.008 | 0.88 | .379 |
| **Random effects** | | | | |
|  | *Variance* | *SD* | | |
| Item-criterion | 0.308 | 0.555 | | |
| Model | 0.032 | 0.178 | | |

**Table A3.** Generalized Linear Mixed Model Predicting Scoring Accuracy from Reasoning Level and Model

|  | *b* | *SE* | *z* | *p* |
|---|---|---|---|---|
| **Fixed effects** | | | | |
| Intercept | 2.263 | 0.210 | 10.75 | < .001 |
| Reasoning (linear) | 0.145 | 0.036 | 4.06 | < .001 |
| Reasoning (quadratic) | 0.008 | 0.035 | 0.22 | .823 |
| Reasoning (cubic) | −0.069 | 0.034 | −2.04 | .041 |
| Gemini 3.1 Pro Preview | 0.260 | 0.060 | 4.33 | < .001 |
| GPT-5.4 | 0.079 | 0.052 | 1.53 | .126 |
| GPT-5.4 Mini | 0.084 | 0.052 | 1.61 | .108 |
| GPT-5.4 Nano | 0.095 | 0.052 | 1.82 | .069 |
| **Random effects** | | | | |
|  | *Variance* | *SD* | | |
| Item-criterion | 0.300 | 0.548 | | |

# References

Bates, D., Maechler, M., Bolker, B., & Walker, S. (2015). Fitting linear mixed-effects models using lme4. *Journal of Statistical Software, 67*(1), 1–48. https://doi.org/10.18637/jss.v067.i01

Burrows, S., Gurevych, I., & Stein, B. (2015). The eras and trends of automatic short answer grading. *International Journal of Artificial Intelligence in Education, 25*(1), 60–117. https://doi.org/10.1007/s40593-014-0026-8

DiCerbo, K. (2025, October). *How generative AI is transforming student assessment at Khan Academy*. Khan Academy Blog. https://blog.khanacademy.org/how-generative-ai-is-transforming-student-assessment-at-khan-academy/

Frohn, S., Burleigh, T., & Chen, J. (2025). Automated scoring of short answer questions with large language models: Impacts of model, item, and rubric design. In *Artificial Intelligence in Education, Lecture Notes in Artificial Intelligence* (Vol. VI, pp. 44–51). https://doi.org/10.1007/978-3-031-98465-5_6

Google. (2026). *Gemini developer API pricing*. Google AI for Developers. https://ai.google.dev/gemini-api/docs/pricing

Henkel, O., Hills, L., Boxer, A., Roberts, B., & Levonian, Z. (2024). Can large language models make the grade? An empirical study evaluating LLMs ability to mark short answer questions in K-12 education. In *Proceedings of the Eleventh ACM Conference on Learning @ Scale* (pp. 300–304). https://doi.org/10.1145/3657604.3664693

Lee, G.-G., Latif, E., Wu, X., Liu, N., & Zhai, X. (2024). Applying large language models and chain-of-thought for automatic scoring. *Computers and Education: Artificial Intelligence, 6*, 100213. https://doi.org/10.1016/j.caeai.2024.100213

OpenAI. (2024). *Learning to reason with LLMs*. OpenAI Research Blog. https://openai.com/research/learning-to-reason-with-llms

OpenAI. (2025a). *OpenAI o3-mini*. https://openai.com/index/openai-o3-mini/

OpenAI. (2025b). *Introducing OpenAI o3 and o4-mini*. https://openai.com/index/introducing-o3-and-o4-mini/

OpenAI. (2026). *Pricing*. OpenAI Developer Platform. https://developers.openai.com/api/docs/pricing

Pack, A., Barrett, A., & Escalante, J. (2024). Large language models and automated essay scoring of English language learner writing: Insights into validity and reliability. *Computers and Education: Artificial Intelligence, 6*, 100234. https://doi.org/10.1016/j.caeai.2024.100234

Sui, Y., Chuang, Y.-N., Wang, G., Zhang, J., Zhang, T., Yuan, J., Liu, H., Wen, A., Zhong, S., Zou, N., Chen, H., & Hu, X. (2026). *Stop overthinking: A survey on efficient reasoning for large language models* (arXiv:2503.16419). https://doi.org/10.48550/arXiv.2503.16419

Wang, X., Wei, J., Schuurmans, D., Le, Q., Chi, E., Narang, S., Chowdhery, A., & Zhou, D. (2023). *Self-consistency improves chain of thought reasoning in language models* (arXiv:2203.11171). https://doi.org/10.48550/arXiv.2203.11171

Wei, J., Wang, X., Schuurmans, D., Bosma, M., Ichter, B., Xia, F., Chi, E., Le, Q., & Zhou, D. (2023). *Chain-of-thought prompting elicits reasoning in large language models* (arXiv:2201.11903). https://doi.org/10.48550/arXiv.2201.11903

Wickham, H., Cheng, J., Jacobs, A., Aden-Buie, G., & Schloerke, B. (2025). *ellmer: Chat with large language models* (R package version 0.3.2). https://doi.org/10.32614/CRAN.package.ellmer

Xue, M., Xiao, X., Liu, Y., & Wilson, M. (2026). On the consistency of automatic scoring with large language models. *Educational and Psychological Measurement*. https://doi.org/10.1177/00131644261418138

Yoshida, L. (2025). Are the reasoning models good at automated essay scoring? In *Findings of the Association for Computational Linguistics: EMNLP 2025* (pp. 8388–8394).